# Transition metal impurity effect on charge and spin density in iron: *Ab initio* calculations and comparison with Mössbauer data

A. Błachowski[1*] and U. D. Wdowik[2]

[1] Mössbauer Spectroscopy Division, Institute of Physics, Pedagogical University
*PL-30-084 Kraków, ul. Podchorążych 2, Poland*

[2] Institute of Technology, Pedagogical University
*PL-30-084 Kraków, ul. Podchorążych 2, Poland*

[*] Corresponding author: sfblacho@cyf-kr.edu.pl

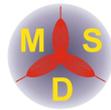

**Abstract**

Density functional theory was applied to study influence of the isolated impurity located on the regular site of the α-Fe crystal on the charge and spin density (hyperfine interactions) on the iron nucleus. Calculations were performed by using both pseudopotential and the full potential methods. The scalar relativistic approximation was applied. Perturbations of the charge and spin density on iron were calculated for all d impurities soluble in iron and additionally for Ga impurity. It was found that impurities have measurable effect on the iron charge and spin density up to the second or third coordination shell depending on the impurity. Hyperfine parameters of iron adjacent to the impurity are affected by two intermixed physical mechanisms, i.e., the volume mismatch due to the impurity and electron band mixing caused by the electronic configuration of the impurity outer shells. Some correlations between *ab initio* calculations and Mössbauer experimental results are discussed. A table is provided with the parameters allowing calculate Mössbauer spectrum of the binary iron alloy with d impurity or Ga. On the other hand, provided parameters allow extraction from the Mössbauer data information about impurity concentration and eventual order.

## 1. Introduction

Iron alloys with transition metals are technologically important materials, as they are base of many steels. The Mössbauer spectroscopy is a common tool to search for the microscopic properties of such alloys. Hence, it is important to learn about d impurity influence on the iron hyperfine parameters versus distance between impurity and iron atom.

There are plenty binary alloys basing on the α-Fe BCC structure with many diluted impurities dissolved randomly on the regular iron sites [1]. The majority of the impurities dissolving in iron belong to the block d – transition metals. These impurities have influence on the iron hyperfine magnetic field and on the isomer shift observed by $^{57}$Fe Mössbauer spectroscopy. Hence, they modify electron spin and charge densities on the adjacent iron nuclei. Above perturbations are additive algebraically, and they depend on the coordination shell of the iron with the impurity replacing iron atom. Perturbations due to single impurity could be seen to the second or third coordination shell depending on the impurity [2-4].

This work reports *ab initio* investigations of impurity effect on the charge and spin densities on the iron nucleus. First principles studies were performed for impurities located at some definitive coordination shell of the iron atom in order to make some comparison with the available experimental Mössbauer data.

## 2. Computational details

Calculations were performed for all d atoms (transition metals) soluble in α-Fe and additionally for gallium impurity. Pure α-Fe was also calculated in order to have reference data. All calculations are carried out using the spin-polarized density functional theory (DFT). The isolated impurity was incorporated substitutionally into the center of the supercell composed of the Fe atoms arranged on the BCC lattice. Symmetry constraints were imposed on the supercell containing 128 atoms. This enabled to calculate hyperfine parameters for all Fe atoms which belong to the subsequent eleven shells around the impurity as well as for 13$^{th}$ and 17$^{th}$ shell (see Table 1). Calculations were divided into two steps. In the first step the pseudopotential method with the generalized gradient approximation (GGA) parameterized by Perdew-Burke-Ernzerhof method (PBE) as implemented in the VASP code [5] was used to optimize geometry and atomic positions of the supercell. Atoms were represented by the projector-augmented wave pseudo potentials (PAWs) provided by VASP. A plane-wave expansion up to 360 eV was applied. The Brillouin zone of each supercell with impurity was sampled using the 3x3x3 *k*-point mesh generated by the Monkhorst-Pack scheme. The lattice vectors of the supercell were frozen at the GGA optimized value and the atomic positions were relaxed until the forces acting on all atoms of the supercell were smaller than 0.001 eV/nm. The total energy was converged down to 0.1 meV/supercell. The second step was concerned with the calculation of the hyperfine parameters on Fe nucleus using relaxed atomic positions from the first step. Hyperfine parameters were calculated using the full potential all electron plane wave method (FLAPW) as implemented in the WIEN2k code [6]. The scalar relativistic approximation was used as the spin-orbit interactions are irrelevant for binary alloys of iron [7]. These calculations were carried out using the same exchange-correlation approximation and 4 *k*-points in the irreducible part of the Brillouin zone. The muffin-tin radii of particular atoms were set to conform to the criterion of the almost touching spheres. The cut-off energy expressed as the product of the muffin-tin radii and the maximum plane wave vector was equal to 7 and the largest reciprocal vector for the charge Fourier transfer amounted to 12. The total energy convergence of the order of 0.01 mRy/supercell was achieved. Electronic states lying more than 7 Ry below Fermi level were treated as the core states. Finite nuclear dimensions were used in the calculations in the same manner as in Ref. [8]. A nuclear radius of the $^{57}$Fe nucleus (in the ground state) was used for Fe atoms as the electron density on the particular nucleus depends on the nuclear radius of this nucleus and it is practically independent of the radii of the remaining nuclei. Hyperfine parameters were calculated for all iron containing coordination shells around impurity with various radii up to the 11$^{th}$ shell and for 13$^{th}$ and 17$^{th}$ shells.

**Table 1**

Coordination shells around impurity located in the zero-th shell ordered by distance from the impurity. The symbol $N_{cell}$ denotes number of atoms which belong to the particular shell and to the 128-atomic supercell with impurity in the center. The symbol $N_{shell}$ stands for the number of all atoms in the particular shell (distance) in the BCC lattice, while the symbol $r/a$ denotes relative distance (normalized by the lattice constant $a$) from the impurity in the unrelaxed ideal BCC lattice. Relaxation of the radial distances do not change neither the order of shells nor the number of atoms in particular shell.

| Shell No | $N_{cell}$ | $N_{shell}$ | $r/a$ |
|---|---|---|---|
| 0 | 1 | 1 | 0 |
| 1 | 8 | 8 | 0.8660 |
| 2 | 6 | 6 | 1 |
| 3 | 12 | 12 | 1.4142 |
| 4 | 24 | 24 | 1.6583 |
| 5 | 8 | 8 | 1.7321 |
| 6 | 3 | 6 | 2 |
| 7 | 24 | 24 | 2.1795 |
| 8 | 12 | 24 | 2.2361 |
| 9 | 12 | 24 | 2.4495 |
| 10 | 8 | 32 | 2.5981 |
| 11 | 3 | 12 | 2.8284 |
| 13 | 6 | 30 | 3 |
| 17 | 1 | 8 | 3.4641 |

## 3. Results

Calculated variations between distance from the impurity to the nearest neighbor iron atom $\Delta r_1$ relative to the distance between nearest iron atoms in the pure iron are shown versus the number of the impurity outer shell electrons $N_e$ in the upper part of the Figure 1. Experimental variation of the lattice constant $a$ with the impurity concentration $c$, i.e., $da/dc$ versus number $N_e$ is shown in the lower part of Figure 1. Experimental data were obtained by means of the powder X-ray diffraction method at room temperature and they are taken from the literature [9-11]. Parameters $\Delta r_1$ and $da/dc$ generally increase while departing from $N_e = 8$, i.e., the number of outer electrons in iron. They are smaller for 3d elements in comparison with the respective $N_e$ of 4d or 5d elements. There is practically no difference between corresponding 4d and 5d elements. Relaxation of the atomic distances is an important step in order to have reliable basis for further calculations of the hyperfine parameters as the volume effect strongly influences hyperfine interactions [12].

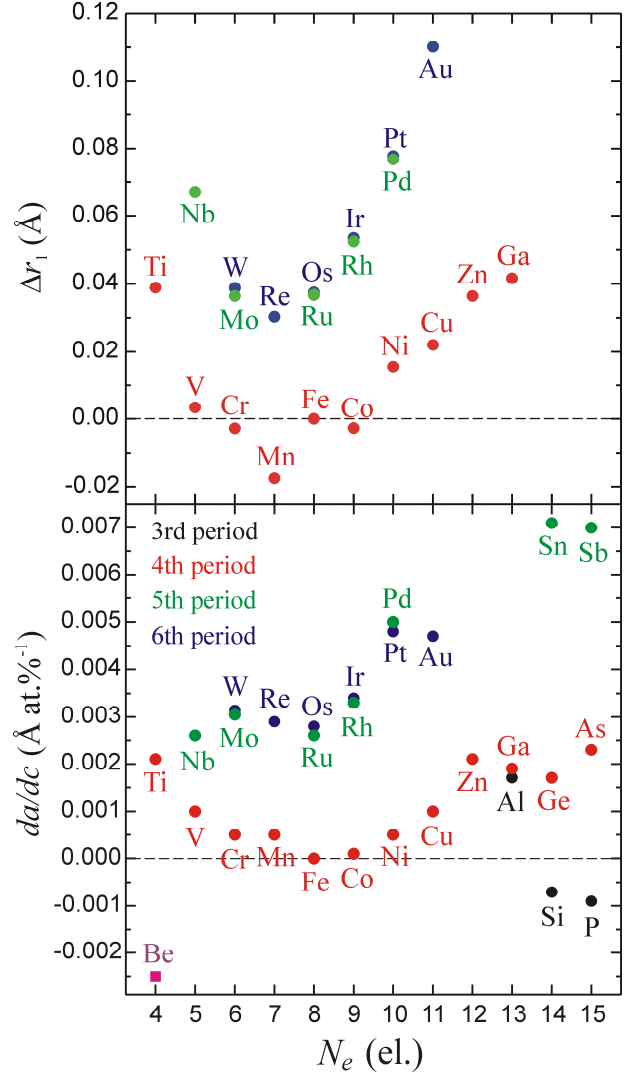

**Figure 1** Calculated difference in distance between impurity and the nearest iron atom and distance between adjacent iron atoms in the pure iron $\Delta r_1$ plotted versus the number of the outer shell electrons of the impurity $N_e$ (upper part). Corresponding experimental variation of the lattice constant with the impurity concentration $da/dc$ [9-11] plotted versus $N_e$ (lower part).

Perturbations of the electron (charge) $\Delta \rho_s$ and electron spin $\Delta B_s$ density were calculated as follows. The average value of the electron (charge) $\rho_0$ or spin $B_0$ density was calculated for iron atom located from 7$^{th}$ till 13$^{th}$ coordination shell excluding 12$^{th}$ shell. It is clear that iron atom in the 7$^{th}$ or more distant coordination shell from the impurity behaves like iron atom in the pure iron (see Figure 2). It means that impurity located in the 7$^{th}$ or more distant coordination shell is invisible to the iron atom. This average value was used to set reference level of the respective perturbation $\Delta \rho_s$ or $\Delta B_s$. The index $s$ enumerates subsequent coordination shells around impurity. The electron spin densities are recalculated to the hyperfine fields acting on the iron nucleus [6]. Figure 2 shows the electron density perturbation $\Delta \rho$ and the electron spin density perturbation $\Delta B$ versus relaxed distance from the impurity $r$. Relaxation of the distance between atoms is essential to obtain reliable electronic structure particularly for impurities having different core radius from the iron core.

Points of Figure 2 correspond to subsequent coordination shells around impurity up to the 9th shell. Significant perturbations are seen generally to the third coordination shell unless they are very small and dominated by the numerical fluctuations. Perturbations $\Delta\rho_s$ and $\Delta B_s$ are listed in Table 2 up to the third coordination shell.

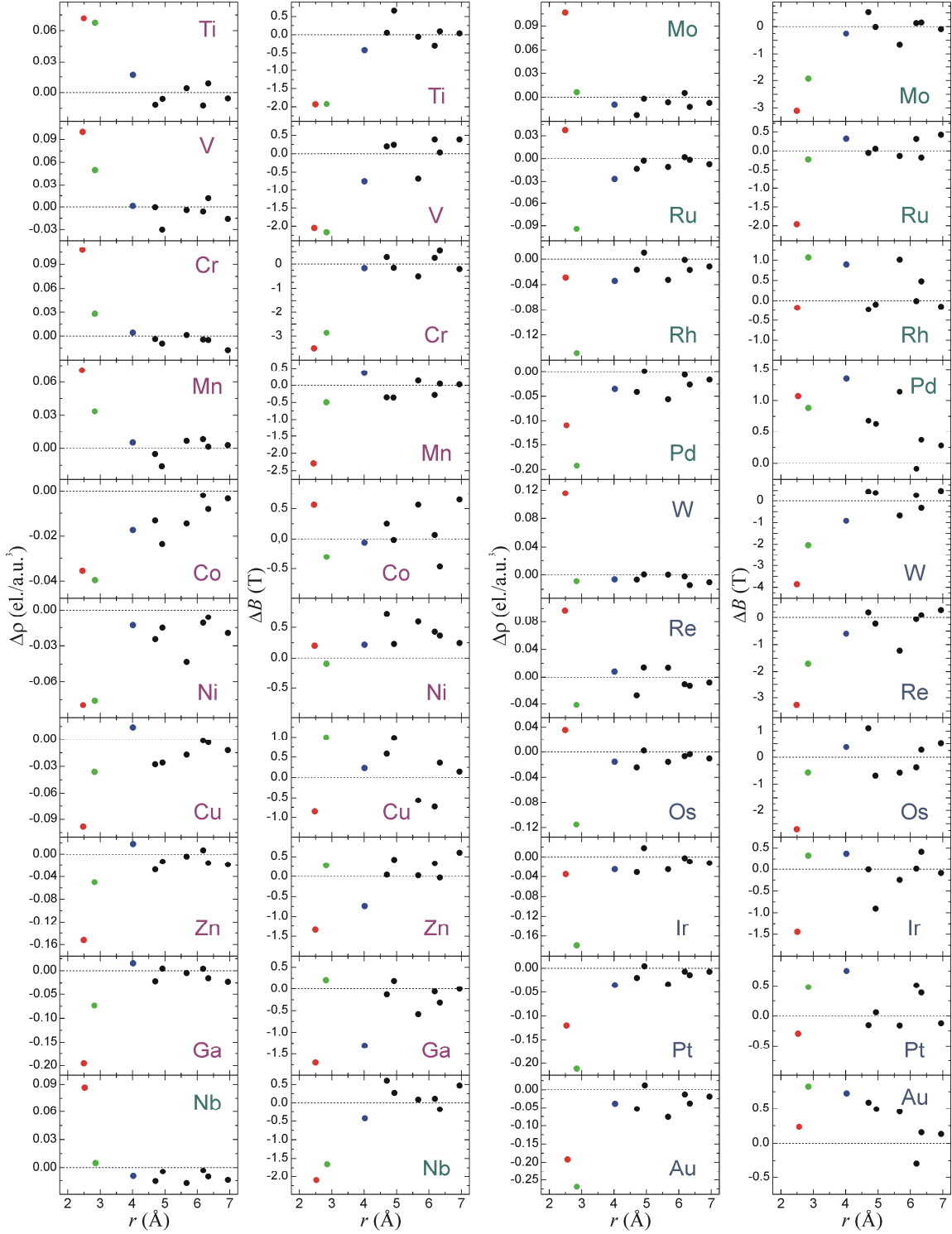

**Figure 2** Calculated electron density perturbations $\Delta\rho$ and spin density perturbations $\Delta B$ plotted versus relaxed distance $r$ from the impurity for various impurities. The zero level has been determined as described within text. Color distinguishes three first coordination shells.

**Table 2**

Perturbations of the electron density $\Delta\rho_s$ and electron spin density $\Delta B_s$ on the iron nucleus generated by the single impurity located in subsequent iron coordination shells. Corresponding perturbation of the isomer shift amounts to $\Delta S_s = \alpha \Delta\rho_s$ with $\alpha = -0.291\,\mathrm{mm\,s^{-1}\,a.u.^3\,el.^{-1}}$ for the 14.41-keV resonant transition in $^{57}$Fe [8]. Perturbation $\Delta B_s$ directly applies to the iron hyperfine magnetic field.

| Impurity | $\Delta\rho_1$ (el./a.u.$^3$) | $\Delta\rho_2$ (el./a.u.$^3$) | $\Delta\rho_3$ (el./a.u.$^3$) | $\Delta B_1$ (T) | $\Delta B_2$ (T) | $\Delta B_3$ (T) |
|---|---|---|---|---|---|---|
| Ti | 0.072 | 0.068 | 0.017 | -1.93 | -1.92 | -0.42 |
| V  | 0.100 | 0.050 | 0.001 | -2.05 | -2.16 | -0.75 |
| Cr | 0.108 | 0.028 | 0.004 | -3.51 | -2.86 | -0.17 |
| Mn | 0.071 | 0.034 | 0.005 | -2.29 | -0.50 | 0.37 |
| Co | -0.036 | -0.040 | -0.017 | 0.56 | -0.30 | -0.06 |
| Ni | -0.080 | -0.076 | -0.013 | 0.20 | -0.09 | 0.22 |
| Cu | -0.098 | -0.036 | 0.014 | -0.85 | 1.00 | 0.23 |
| Zn | -0.152 | -0.050 | 0.018 | -1.34 | 0.28 | -0.74 |
| Ga | -0.195 | -0.073 | 0.016 | -1.70 | 0.21 | -1.30 |
| Nb | 0.087 | 0.005 | -0.010 | -2.10 | -1.65 | -0.43 |
| Mo | 0.107 | 0.006 | -0.009 | -3.10 | -1.93 | -0.26 |
| Ru | 0.037 | -0.094 | -0.027 | -1.96 | -0.22 | 0.32 |
| Rh | -0.029 | -0.149 | -0.034 | -0.18 | 1.07 | 0.89 |
| Pd | -0.109 | -0.192 | -0.035 | 1.07 | 0.88 | 1.35 |
| W  | 0.115 | -0.009 | -0.006 | -3.86 | -2.04 | -0.92 |
| Re | 0.097 | -0.041 | 0.008 | -3.27 | -1.71 | -0.60 |
| Os | 0.036 | -0.115 | -0.015 | -2.70 | -0.57 | 0.38 |
| Ir | -0.035 | -0.179 | -0.025 | -1.45 | 0.33 | 0.37 |
| Pt | -0.120 | -0.211 | -0.035 | -0.29 | 0.48 | 0.75 |
| Au | -0.192 | -0.269 | -0.039 | 0.23 | 0.82 | 0.72 |

The average electron density on the iron nucleus perturbed by impurities located up to the first three coordination shells could be expressed as $\langle\rho\rangle_3 = \rho_0 + c\sum_{s=1}^{3} n_s \Delta\rho_s$ with the symbol $n_s$ denoting number of vortices within the $s-th$ shell ($n_1 = 8$, $n_2 = 6$ and $n_3 = 12$ for the BCC structure). Corresponding average hyperfine field (electron spin density) on the iron nucleus is expressed as $\langle B\rangle_3 = B_0 + c\sum_{s=1}^{3} n_s \Delta B_s$. Figure 3 shows $d\langle\rho\rangle_3/dc$ and $d\langle B\rangle_3/dc$ versus number of the impurity outer shell electrons $N_e$. Lower part of Figure 3 shows correlation between above quantities for various impurities. The average electron density on the iron nucleus generally decreases with the increasing number of the impurity outer shell electrons. The relationship is roughly linear with the stronger effect for 4d and 5d impurities as compared to the 3d impurities. The average spin density versus $N_e$ is more complex. It is roughly linear for 4d and 5d metals increasing with the increasing $N_e$ and it has some oscillatory character for 3d metals. Correlation between average electron and spin density indicates that the electron density is lowered with the increasing spin density. The correlation between these two quantities is the weakest for 3d metals.

A total perturbation of the electron density for impurities within the first three coordination shells could be defined as $\delta\rho_3 = \sqrt{\sum_{s=1}^{3} w_s \Delta\rho_s^2}$ with the shell weight factors $w_s = n_s / \sum_{s'=1}^{3} n_{s'}$

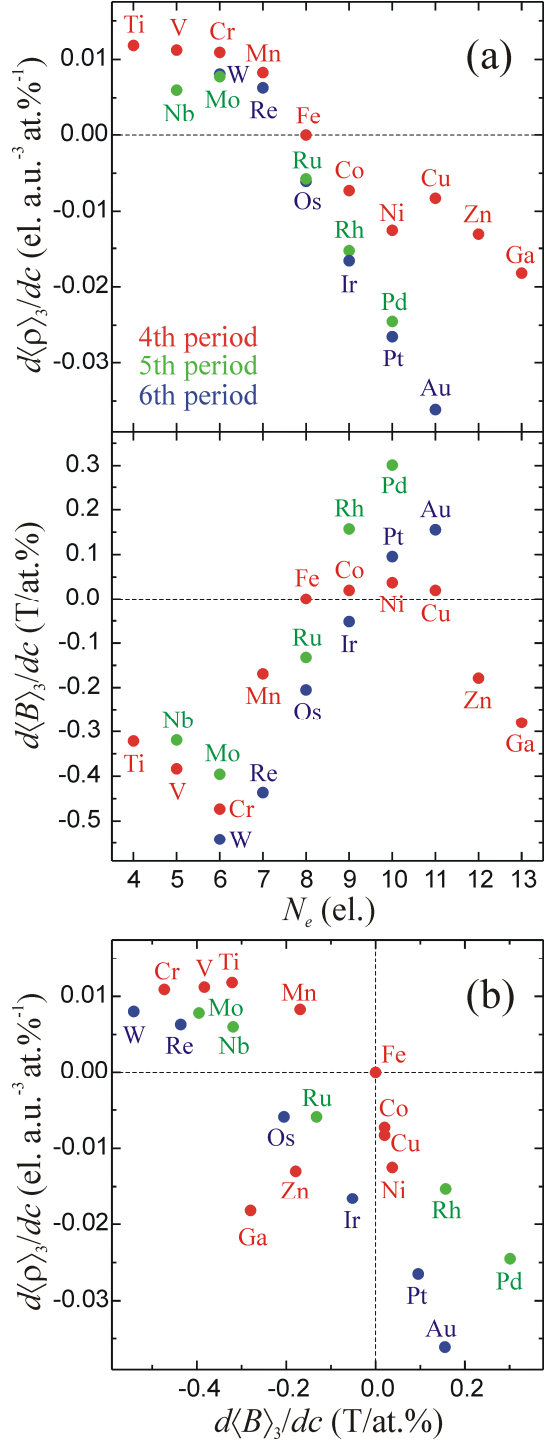

($w_1 = 8/26$, $w_2 = 6/26$ and $w_3 = 12/26$ for the BCC structure). Corresponding total perturbation of the spin density is expressed as $\delta B_3 = \sqrt{\sum_{s=1}^{3} w_s \Delta B_s^2}$. Figure 4 shows $\delta\rho_3$ and $\delta B_3$ versus $N_e$, while the correlation between these two quantities is shown in the lower part of Figure 4 for various impurities. A total electron density perturbation increases with the increasing deviation of the impurity outer shell electrons $N_e$ from the value of the outer shell electrons in iron. The effect is stronger for the positive deviation and for 4d and 5d metals on the positive side. In the case of the total spin density perturbation the effect is stronger for negative deviations in comparison with positive deviations from $N_e = 8$. The main quantum number of the impurity (atomic electron core) seems to be less relevant in comparison with the electron density perturbations as chromium has stronger effect than molybdenum and similar to the effect of tungsten. Correlation between electron density and spin density perturbations show that large perturbations of the electron density correlate in general with small perturbations of the spin density and vice versa.

**Figure 3** (a) Plot of $d\langle\rho\rangle_3/dc$ and $d\langle B\rangle_3/dc$ versus $N_e$ for various impurities, (b) a correlation between $d\langle B\rangle_3/dc$ and $d\langle\rho\rangle_3/dc$.

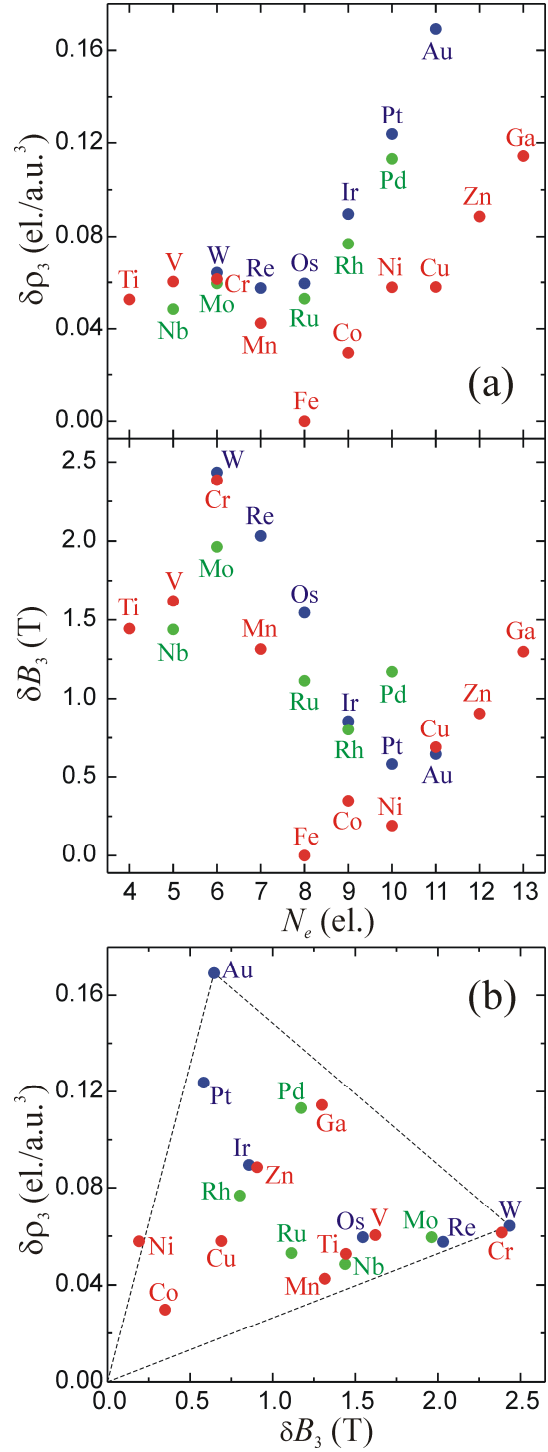

**Figure 4** (a) Plot of $\delta\rho_3$ and $\delta B_3$ versus $N_e$ for various impurities, (b) a correlation between $\delta B_3$ and $\delta\rho_3$ with iron located at the origin.

## 4. Comparison with the experimental Mössbauer data

Results of calculations could be compared with the experimental data obtained by means of the $^{57}$Fe Mössbauer spectroscopy. Hyperfine parameters for the iron binary alloys containing following impurities were taken from the literature: Ti [13], V [14], Cr [15], Mn [2], Co [16],

Ni [2], Cu [17], Zn [17, 18], Ga [19], Nb [20], Mo [21], Ru [22], Rh [23], Pd [24], W [25], Re [26], Os [27], Ir [28], Pt [29, 2] and Au [30] basing on results obtained at room temperature.

Figure 5 shows the experimental $d\langle S\rangle/dc$ and $d\langle B\rangle/dc$ versus calculated $\alpha\,(d\langle\rho\rangle_3/dc)$ and $d\langle B\rangle_3/dc$, respectively. The symbol $\langle S\rangle$ stands for the experimental average isomer (total) shift, the symbol $\langle B\rangle$ for the corresponding experimental magnetic hyperfine field on iron, and the symbol $\alpha = -0.291\,\text{mm s}^{-1}\,\text{a.u.}^3\,\text{el.}^{-1}$ denotes isomer shift calibration constant [8]. A correlation seems reasonable in both cases giving some confidence to the accuracy of calculations. Figure 6 shows experimental total perturbation of the isomer (total) shift $\delta S_E = \sqrt{\sum_{s=1}^{\sigma} w_s^{(\sigma)} \Delta S_s}$ plotted versus calculated $\alpha\,\delta\rho_3$. The shell weight factor is defined as $w_s^{(\sigma)} = n_s / \sum_{s'=1}^{\sigma} n_{s'}$. The symbol $\Delta S_s$ stands for the isomer (total) shift perturbation caused by the single impurity in the $s-th$ shell. Similar correlation between experimental total perturbation of the hyperfine field $\delta B_E = \sqrt{\sum_{s=1}^{\sigma} w_s^{(\sigma)} \Delta B_s}$ and calculated perturbation $\delta B_3$ is plotted in the lower part of Figure 6. The number of subsequent shells taken into account during evaluation of the experimental Mössbauer data amounts to $\sigma = 3$ for Ga, Mo, Ru, Rh, Pd and Ir. For remaining impurities one has $\sigma = 2$. Generally correlation is not very bad, but some impurities show particularly large deviations from the ideal correlation line. It seems that the reason for that is due to difference in the experimental $S_0^{(E)}$, $B_0^{(E)}$ and/or both of them in comparison with the same parameters in the pure iron. The symbol $S_0^{(E)}$ denotes the isomer (total) shift observed in such situation that around iron atom all shells till the shell $\sigma$ inclusive are filled with iron atoms, while the remainder of the sample has random distribution of impurities with definite concentration. The symbol $B_0^{(E)}$ denotes corresponding hyperfine field for above situation. Hence, for $S_0^{(E)}$ being different from the total shift in the pure iron and/or for $B_0^{(E)}$ being different from the hyperfine magnetic field in the pure iron respective differences are due to the global effect of the impurity on the electron and electron spin density on the iron nucleus. For such situations parameters $S_0^{(E)}$ and/or $B_0^{(E)}$ depend on the impurity concentration – usually in the linear fashion, and the correlation of Figure 6 is strongly perturbed.

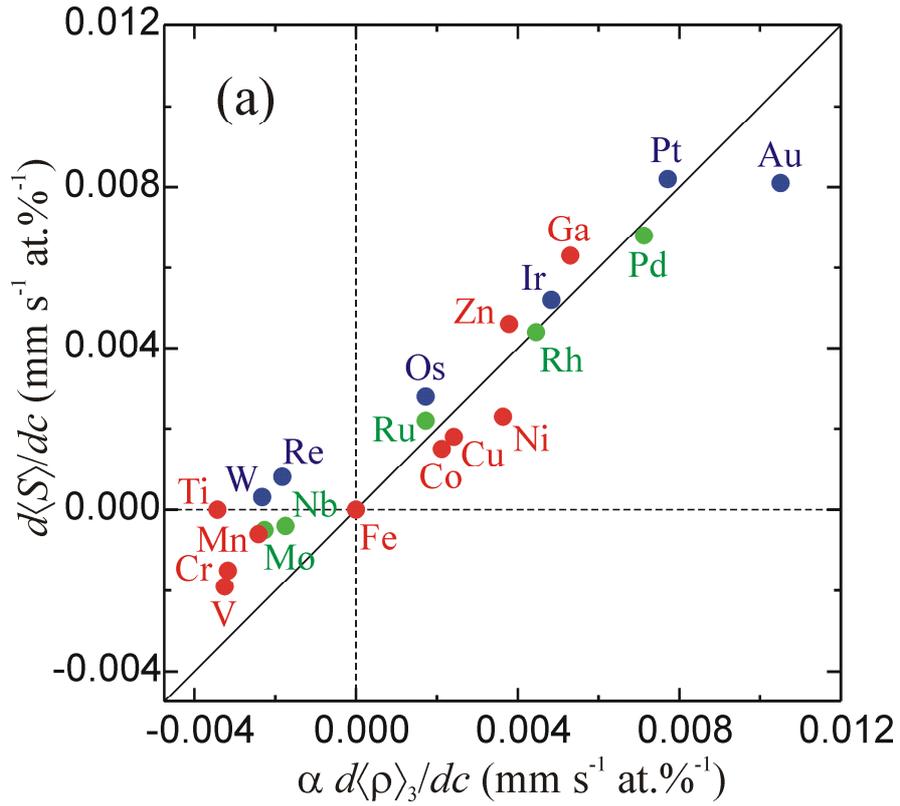

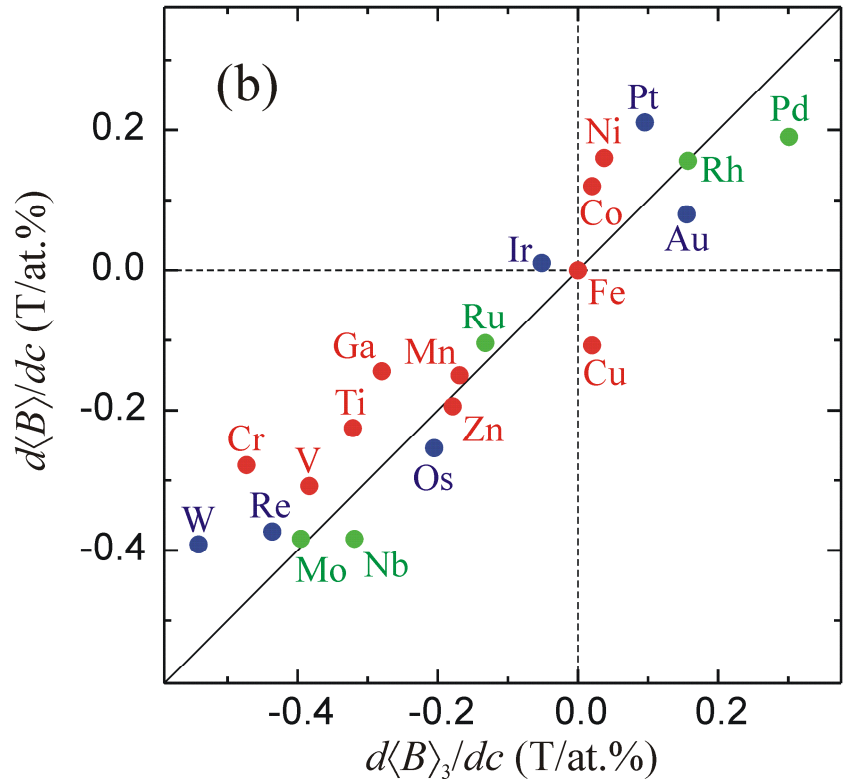

**Figure 5** (a) Correlation between calculated $\alpha (d\langle\rho\rangle_3 / dc)$ and experimental $d\langle S\rangle / dc$, (b) correlation between calculated $d\langle B\rangle_3 / dc$ and experimental $d\langle B\rangle / dc$ for various impurities. Straight lines indicate ideal correlation.

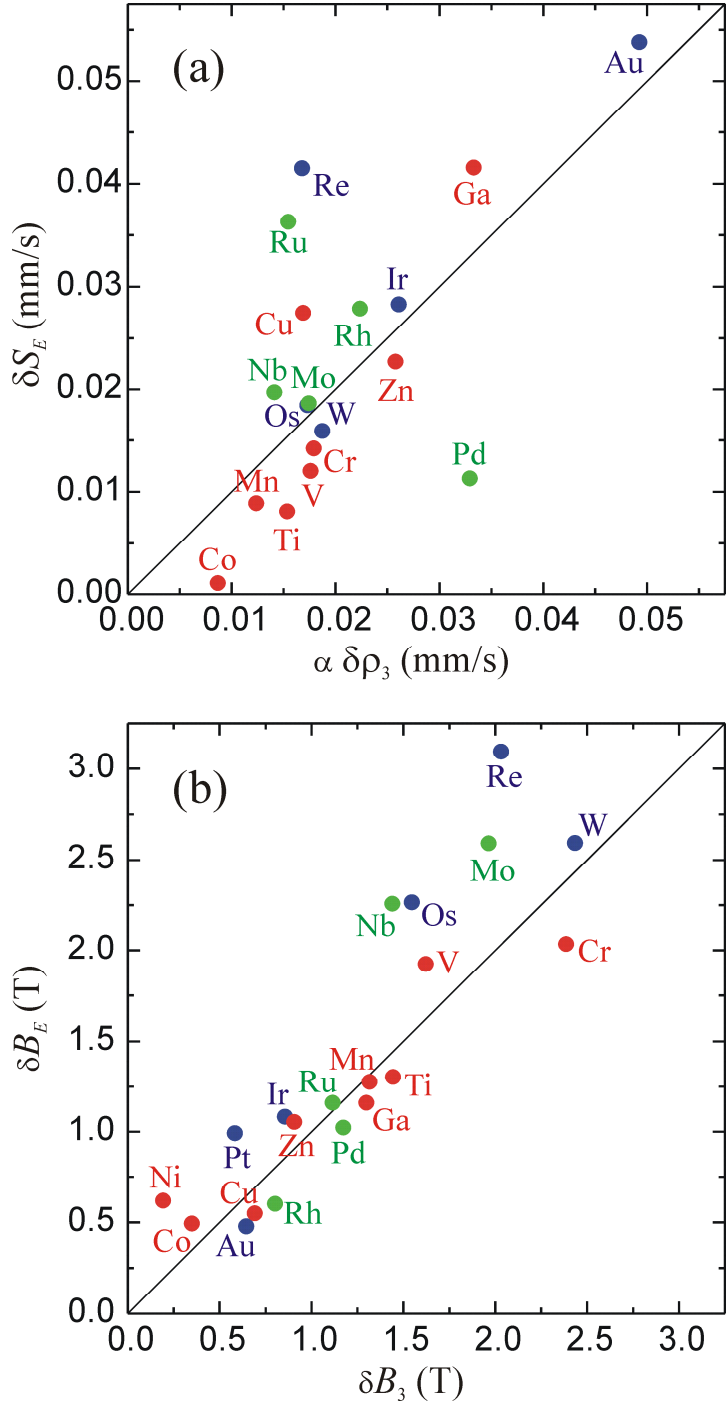

**Figure 6** (a) Correlation between calculated $\alpha\,\delta\rho_3$ and experimental $\delta S_E$, (b) correlation between calculated $\delta B_3$ and experimental $\delta B_E$ for various impurities. Straight lines indicate ideal correlation.

A strong deviation is observed for the electron density (isomer shift) in the case of Pd and Ru, where the parameter $S_0^{(E)}$ depends strongly on the Pd concentration with the positive slope [24], while for Ru one has similar linear dependence with the negative slope [22]. Particular value of the parameters $S_0^{(E)}$ and/or $B_0^{(E)}$ depends on the global properties of the alloy. Hence, these parameters depend on the presence of distant impurities and approach values characteristic of the pure iron with dilution of the alloy.

## 5. Conclusions

Perturbations to the electron density (isomer shift) and the electron spin density (magnetic hyperfine field) on the iron nucleus were calculated for the isolated impurity located at various distances (coordination shells) around the iron atom in the BCC lattice filled with iron atoms otherwise. In fact, calculations are performed for some very diluted and ordered system with the separation between impurities being four relaxed lattice constants along the main crystal axes. Such conditions are imposed by the periodic boundary conditions applied in the calculations. Results of calculations reasonably correlate with the experimental Mössbauer data. On the other hand, experimental situation is somewhat different as one has more or less random alloy with some definite impurity concentration which is usually much higher than the impurity concentration of $1/128 \approx 0.78$ at.% used in the present calculations. It should be mentioned that within the supercell approach calculations of the local perturbations for the random alloy with significant impurity concentration are still limited due to the enormous number of possible and relevant configurations which has to be taken into account. However, as far as the global parameters $S_0^{(E)}$ and/or $B_0^{(E)}$ do not depend on the impurity concentration one can calculate Mössbauer spectra of the BCC binary iron alloys relying on the data of Table 2. On the other hand, provided parameters allow extraction from the Mössbauer data information about impurity concentration and eventual order.

Basically hyperfine parameters of iron adjacent to the impurity are affected by two intermixed physical mechanisms, i.e., the volume mismatch due to the impurity and electron band mixing caused by the electronic configuration of the impurity outer shells. Hence, similarity of the perturbation of the hyperfine parameters due to Cr, Mo and W is caused by similarity of the outermost electronic configuration of these three atoms, while the strong perturbation caused by e.g. Au is mainly due to the volume effect.

It has been found previously [31] that *ab initio* calculated perturbations of the electron density $\Delta S_1$ (see Table 2) are in fair agreement with the same perturbations obtained within cellular atomic model of Miedema and van der Woude [32, 33]. Recent *ab initio* studies of the hyperfine fields in the iron alloys with 3d impurities performed by Rahman *et al.* [7] are in good agreement with our results.

**Acknowledgment**

Interdisciplinary Centre for Mathematical and Computational Modeling (ICM), Warsaw University, Poland is acknowledged for providing the computer facilities under the Grant No. G28-12.